\documentclass[]{JHEP}
\usepackage{epsfig}
\usepackage{amssymb}
\title{Beyond Eikonal Scattering in M(atrix)-Theory}
\author{Robert Helling\thanks{Address from October 2000:
    Humboldt-Universit\"at zu Berlin}\\ Albert-Einstein-Institut\\ MPI f\"ur
  Gravitationsphysik\\ Am M\"uhlenberg 1\\ D-14476 Potsdam\\ Germany}
\abstract{We study the problem of more general kinematics for the finite $N$
  M(atrix)-Model than the simple straight line motion that has been used
  before. This is supposed to be related to momentum transferring processes in
  the dual super-gravity description. We find a negative result for
  classical, perturbative processes and discuss briefly the possibility of
  instianton like quantum mechanical tunneling processes.}

\def\sla#1{\hbox to 0cm{/\hss}#1}
\def\Tr{\hbox{Tr}}
\def\MR{{\mathbb{R}}}
\def\w{\omega}
\def\MM{M(atrix)-Model}

\keywords{M-Theory, D-Branes}
\preprint{AEI-2000-048}

\begin{document}

\section{Introduction}
When the M(atrix)-Model for M-Theory was proposed by Banks, Fischler, Shenker
and Susskind\cite{BFSS} and later in a version for finite $N$ by
Susskind\cite{Su} it was one of the strongest quantitative tests that it was
able to reproduce graviton scattering amplitudes of eleven-dimensional
supergravity. In the sequel, several authors\cite{DKPS, BB, BBPT, K, McA, BHP,
  TvR, OY1,PSW,HKS} were able to generalize the agreement of the famous
$v^4/r^7$ amplitude to higher loop orders, polarization dependent effects and
more particles. Although in the meantime, some of those results have been
given an interpretation in terms of supersymmetric non-renormalization
theorems \cite{PSS1, NP1} and it was shown that the agreement does not persist
if one includes quantum corrections on the sugra side\cite{HPSW}, to our
understanding the ultimate fate of the M(atrix) conjecture, whether it is
applicable at least in a certain regime or the agreement --- especially of the
finite $N$ version that is under computational control --- was a mere
coincidence is not settled, yet.

All the calculations mentioned above have a common kinematical restriction:
They compute quantum fluctuations around a classical solution that represents
{\em free} particles, implying that one has always only studied the limit of
vanishing momentum transfer. In more technical terms, what has been actually
computed is a phase shift in the eikonal regime. On the sugra side, this
either corresponds to a source-probe approximation or, in the more general
language of Feynman graphs, to T-channel scattering\cite{PSW}, see also the
discussion in \cite{HPSW}.

In beginning of this study, we were trying to get rid of this rather strict
kinematical restriction. But it turned out, at least for finite $N$, that this
is impossible, at least for the classical situation. Rather, one has to resort
to quantum mechanical tunneling processes. This means, momentum transfer is
always a non-perturbative process that vanishes in the classical limit.

As the M(atrix)-Model is formulated in the light-cone frame, there are two
different flavors of momentum to both of which our claim applies: First,
there is ordinary momentum in the transversal directions and second, there is
momentum in the light like direction that appears as the the size $N$ of the
matrices, or more specifically as the sizes $N_I$ of the blocks of block
diagonal matrices representing particles with $N_I$ units of light-cone
momentum. Both kinds of momenta are related by a $M^{-i}$ Lorentz boost that is
non-trivial in light-cone coordinates. While the transfer of the latter kind
has long been suspected to be a non-perturbative process there was no reason
to assume the transfer of transversal momentum to be impossible classically.

To render these statement more concrete let us recall the Lagrangian of the
M(atrix)-Model
\begin{equation}
  \label{eq:MMLag}
  L=\Tr\left(\frac 12\dot X^i\dot X^i +\frac 14[X^i,X^j][X_i,X_j]
    -i\bar\psi\dot\psi +\bar\psi\gamma^i[X_i,\psi]\right)
\end{equation}
For the purpose of this investigation we can ignore the fermions. The
potential energy of the bosons 
\begin{displaymath}
  V = -\frac 14[X^i,X^j][X_i,X_j]
\end{displaymath}
is non-negative and vanishes if all the $X^i$ mutually commute. Thus, a
solution to the equations of motion (further on termed ``the trivial'') is that
of diagonal matrices with linear dependence on time:
\begin{displaymath}
  X_{triv}^i(t) = \pmatrix{ r^i_1(t)&&\cr
    &\ddots&\cr
    &&r^i_N(t)\cr
    },\qquad
  r^i_a(t)=b^i_a+v^i_a t
\end{displaymath}
The important observation of \cite{BFSS,W} was to interpret these diagonal
matrix entries as the time-dependent coordinates of $N$ partons (or
D0-particles) in $\MR^9$. For the trivial solution the motion of the partons
is free (at least classically), $\vec b_a$ is the impact parameter of
parton $a$ and $\vec v_a$ is its velocity.
 
The usual procedure is now to employ the background field method by splitting
the quantum field $X^i$ into this classical part and small quantum
fluctuations:
\begin{displaymath}
  X^i(t)=X^i_{triv}(t)+Y^i(t)
\end{displaymath}
The vacuum effective action of the $Y^i$ is then the effective action of $N$
partons with asymptotic states described by the $\vec b_a$ and $\vec
v_a$. Performing this calculation at the one-loop level yields the famous
$v^4/r^7$ potential of 11D supergravity.

It is important to include the fermions in the quantum part of this
computation whereas in the classical background, they only encode
polarizations of the gravitons\cite{K,BHP}. The leading order interaction is
always given by a purely bosonic background. Therefore, for the purpose of our
investigation which aims to generalize the classical background we do not have
to take into account the fermions.

Due to the nature of the trivial solution, it was so far only possible to
compute the quantum effective action for states for which the in and the out
state are identical and differ only by a mere phase. In the supergravity
language this means that no momentum has been transferred and the scattering
has been performed in the T-channel.

In the setting of the background field method, in order to drop this strong
kinematical restriction one would have to find a classical solution with the
appropriate asymptotic behavior and perform the fluctuation analysis in this
background.

As we are aiming at scattering processes we are not interested in bound
solutions that to not approach infinity for asymptotic times. Now we can
formulate the problem investigated in this paper in a gauge independent way:
We are looking for solutions of the bosonic equations of motion
\begin{displaymath}
  \ddot X^i = -[X^j,[X^i,X^j]]
\end{displaymath}
that escape to infinity, i.e.
\begin{displaymath}
  R^2(t) := \Tr \left(X^i(t) X^i(t)\right),\qquad 
  \lim_{t\to\pm\infty}R^2(t) = \infty. 
\end{displaymath}
Due to the nature of the potential, impact parameters and velocities will again
be well defined for asymptotic times. 

As it will turn out, only the trivial solution which is characterized by the
exact vanishing of the potential energy for all times fulfills these
requirements. Thus, it is impossible to transfer momentum of any kind in the
finite $N$ matrix model, at least in perturbative processes. This is
surprising as one might well have imagined many more, possibly very
complicated, classical solutions that asymptotically escape along the valleys
of the potential.

It is important to point out that this result only applies to the finite $N$
version of the M(atrix)-Model. For infinite $N$, one can T-dualize one
direction and obtain M(atrix)-String-Theory of \cite{DVV}. There, the existence
of non-trivial scattering solutions is well known\cite{GHV,AF1,AF2}.

In the following section, we will present our argument for a highly simplified
model. This will be done in some detail, as in section three, the discussion
of the full model can be reduced to this simplified case. In section four, we
will mention some results for the Wick rotated model, there we will find
instanton like processes that are classically forbidden. In a final section, we
collect our conclusions.

\section{The Toy Model}
In \cite{dWLN} a simple toy model that mimics the quartic interactions found
in the M(atrix)-Model was introduced. Here in this section, we will develop
the main strategy we are going to follow. This is done in quite some detail
not only for pedagogical reasons but later it will turn out that the more
general case can in fact be reduced to the discussion of the toy model.

In the toy model, there are only two real degrees of freedom denoted by $x$ and
$y$ with a Hamilton function
\begin{displaymath}
H=\frac 12 \left(\dot x^2+\dot y^2+x^2y^2\right)
\end{displaymath}
Although this is symmetric in $x$ and $y$ one should think of $x$ representing
a coordinate along the valley while $y$ represents a coordinate transversal to
the valley.

This toy model can be understood as a truncation of the M(atrix)-Model as
follows: Let $e_1, e_2\in u(N)$ be two constant matrices that obey
\begin{displaymath}
  \Tr (e_A, e_B) = \delta_{AB},\qquad [e_1,e_2]\ne 0.
\end{displaymath}
Then, with the ansatz
\begin{displaymath}
  X^1(t)=x(t)e_1, \qquad X^2(t)=y(t)e_2,\qquad X^i=0\hbox{ for }i>1
\end{displaymath}
we recover the toy model. By numerical investigations in the context of QCD it
has been known for a long time that this model generically exhibits chaotic
behavior. Thus, we cannot hope to find exact solutions except in very
singular initial conditions. Nevertheless, we will be able to make statement
about properties of the general solution.

Just as in the full model, the potential
\begin{displaymath}
  V=\frac 12 x^2y^2
\end{displaymath}
has valleys that reach infinity and become narrower away from the ``stadium''.
More precisely, transversal to the valley, the potential is quadratic with
curvature proportional to the distance from the origin.
\EPSFIGURE{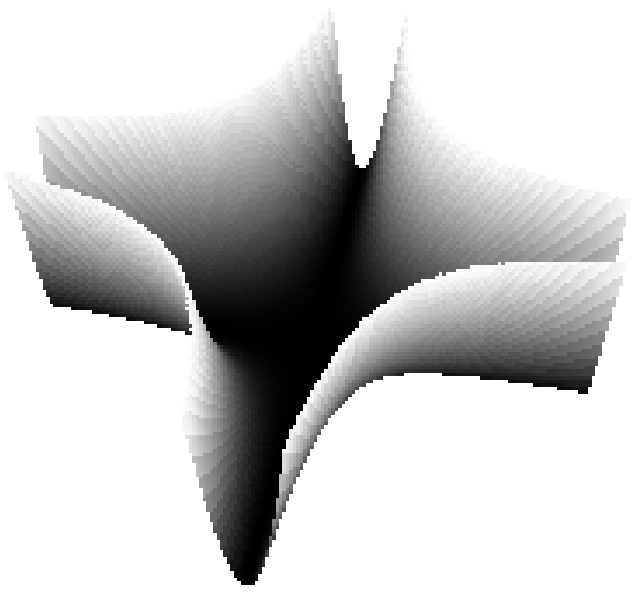}{The potential of the toy-model}
This can also be seen from the equations of motion
\begin{displaymath}
  \ddot x=-xy^2,\qquad \ddot y=-yx^2.
\end{displaymath}
As in the full model, there is a trivial solution if both sides of the
equations vanish identically:
\begin{displaymath}
  x(t) = b + vt,\qquad y(t)=0\qquad\hbox{or }x\leftrightarrow y,
\end{displaymath}
as the ansatz above encompasses the trivial solution of the M(atrix)-Model. As
for the full model, we ask if there are further scattering solutions in the
sense that
\begin{displaymath}
  R(t)^2 = x^2+y^2
\end{displaymath}
goes to infinity for early and late times. This limit is meant in the usual
sense that for each $R_0$ there is a $T$ such that $R(t)^2>R_0$ if $|t|>T$. We
are not interested in solutions that enter the valleys but always return to
the ``stadium'' around the origin after every such excursion, although $R(t)$
might not be bounded for such solutions.

We can use the $x\leftrightarrow \pm y$ symmetry to assume without loss of
generality that the solution escapes along the positive $x$-axis for late
times. From the equation of motion, we see that also $\dot x$ has to be
positive after $T$ because otherwise the velocity will stay negative until
$x=0$ and we are back to the stadium again. As we have argued above, we expect
a motion that is mainly directed along the $x$-axis but with small
oscillations in $y$ that are bound by the valleys. There are two possible
scenarios: Either these oscillations will get smaller and smaller as the
valleys are getting narrower and narrower or the oscillations are so strong
that, eventually, the component of the gradient of the potential in negative
$x$ direction off the bottom of the valley will stop the motion in the
$x$-direction and force the particle to return to the stadium.

Numerical evolutions of the equations of motion indicate that the latter
behavior is generic but we would like to investigate if there can be
exceptions other than the trivial $y=0$ solution we have given above.
\EPSFIGURE{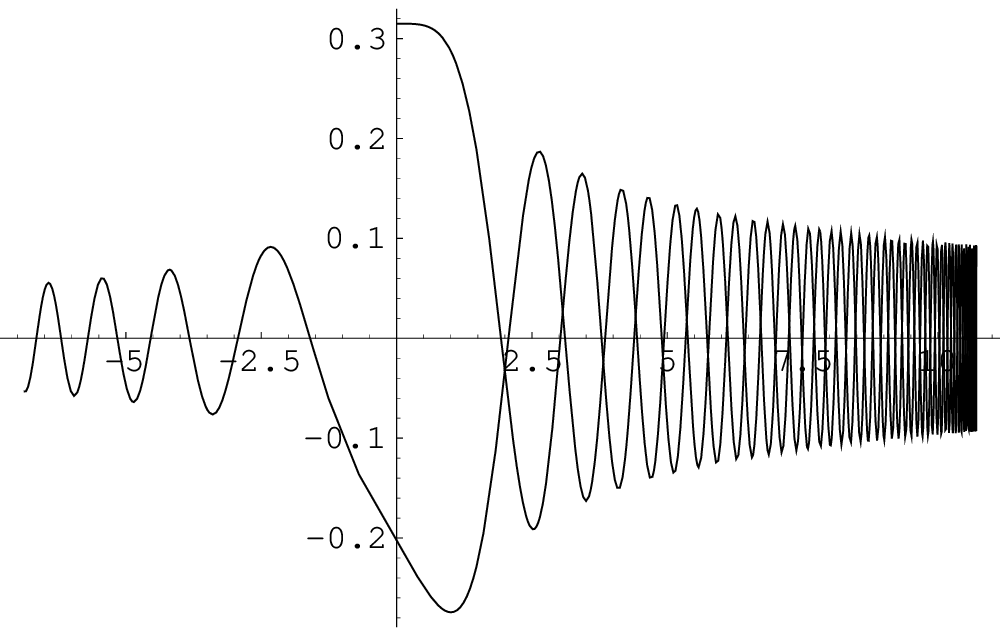}{A numerical example: The solution of the toy model returns
  to the stadium}

It is instructive to split the energy into two contributions coming from 
the motions in the $x$ and $y$ direction:
\begin{displaymath}
  E=E_x+E_y,\qquad E_x=\frac 12 \dot x^2,\qquad E_y=\frac 12\left( \dot
    y^2+x^2y^2\right)
\end{displaymath}
Taking the time derivative, 
\begin{displaymath}
  \dot E_y=-\dot E_x=-\ddot x\dot x=x\dot xy^2\ge 0
\end{displaymath}
where we used the assumptions $x>0$ and $\dot x>0$ from above. We see that
energy is transferred constantly from the $x$ motion to the oscillations in
the $y$ direction. Thus, if we took the energy $E_y$ to be constant we would
{\em under}estimate the oscillations.

Let us recall one fact about the harmonic oscillator: It follows from the
virial theorem $\langle E_{\rm kin}\rangle = \langle E_{pot}\rangle$ that the
time average of the square of the oscillating variable $\langle\phi^2\rangle$
is given by
\begin{equation}
\label{eq:msharm}
  \langle\phi^2\rangle={E\over \w^2}
\end{equation}
in terms of the energy $E$ and the frequency $\w$. Since $x$, the frequency of
the oscillations in $y$, is assumed to go to infinity, $\frac 1x$, the
timescale of the oscillations, is going to zero. Thus, for late times, the
variation of the frequency during one period of the oscillation becomes
smaller and smaller. Therefore, in the equation of motion for $x$, we can
replace the effect of the oscillations by the average over one period and use
\ref{eq:msharm}:
\begin{displaymath}
  \ddot x=-xy^2\approx -x\langle y^2\rangle=-x{E_y\over x^2}=-{E_y\over x}
\end{displaymath}
This force on $x$ can be described by an effective potential as
\begin{displaymath}
  V_{\rm eff}(x)= E_y\log x.
\end{displaymath}
\EPSFIGURE{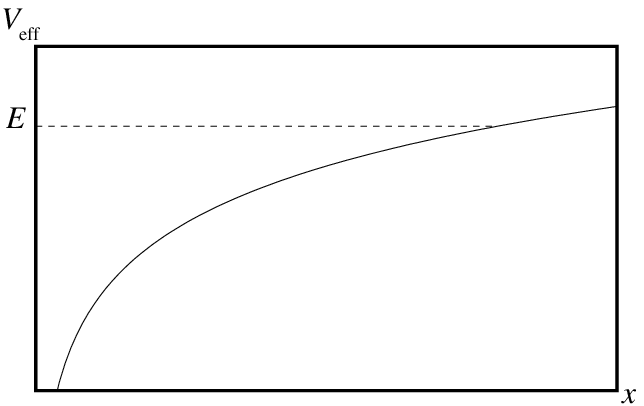}{The effective potential}

As the logarithm grows without bound, the motion in $x$ will hit a potential
barrier no matter how big the total energy $E$ is. The only exception would
be that $E_y=0$ but this is again the trivial solution without
oscillations. Therefore, we do not expect to find any other solutions that
escape to infinity.

One might worry that the above reasoning using the adiabatic time averaging
might be not justified. Therefore, we will give a more formal proof of the
non-existence of scattering solutions, next. To this end, we will proceed
along the lines of the previous heuristic argument, but without using any a
priori knowledge about the solution. Therefore, we cannot employ the
expressions for the harmonic oscillator in the $y$ direction.

Our strategy will be to assume that there is a solution $(x(t),y(t))$ with
$E_y>0$ for which $x(t)$ goes to infinity as $t\to\infty$, and to show that
this assumption leads to a contradiction. Namely, we will show that under
these assumptions
\begin{displaymath}
  \int_0^\infty dt\, \ddot x=-\infty.
\end{displaymath}
This implies that any velocity in the $x$ direction will be stopped and $x$
will eventually become negative again. Thus, only the trivial solutions
mentioned above will escape to infinity. Of course, as the model is invariant
under time reversal, this also means that every solution that comes from
$x=-\infty$ in the past has to be trivial.

To begin with, let us recall that we can assume $x$ to be arbitrarily large
and that $\dot x$ is positive. Furthermore, $E_y$ is strictly increasing in
time and since the total energy is conserved, all velocities and $y$ are
bounded by constants determined by the initial conditions.

For some $t_0$, label the $y$-axis such that $y(t_0)\le 0$. The first thing to
notice is that the retraction force is bounded by the force of the harmonic
oscillator of the ``momentary'' frequency $\w_0=x(t_0)$ as long as $y$ is
negative. But as a harmonic oscillator returns to $0$ within the next time
interval of length $\pi/\w_0$, the $y$ motion has to cross the  $x$ axis
within this interval of time , too. Let us call this moment of $y=0$ the time
$t=0$.

Comparing with the harmonic oscillator once more, we can conclude that the
motion again is bounded from above by that of the harmonic oscillator and that
$y$ will eventually return to $0$ at some moment $0<t_1<\pi/\w_0$.
\EPSFIGURE{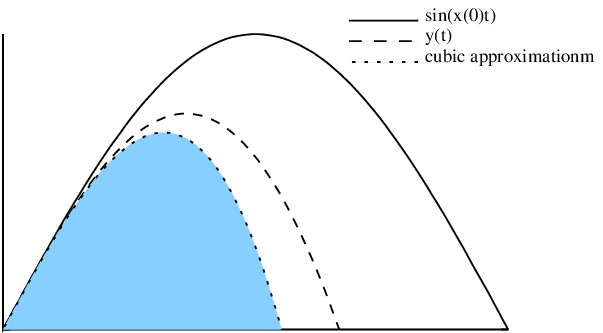}{Upper
and lower bounds on $y(t)$}

Now, we are going to estimate the retarding effect of the oscillations in $y$
on the motion in $x$. Since we are going to show that the retardation is
strong enough to stop the motion in $x$ we have to give a {\em lower} bound on
$y$. At $t=0$, the velocity $\dot y$ is positive. Any trial function $\tilde
y$ with $\tilde y(0)=0$, $\dot{\tilde y}(0)=\dot y(0)$, and $\ddot{\tilde
y}(t)<\ddot y(t)$ will have this property. As the kinetic energy is bounded
from above, so is the velocity. Thus, by assuming $x(0)$ large enough, we know
that $x(t)$ cannot double over one period. Hence, with the help of the
inequality
$$-\ddot y(t)\le 4t \dot y(0)x(0)^2,$$
which holds because assuming the velocity in $y$ to be constant is an
overestimate, we can underestimate $y(t)$ by truncating its Taylor series at
the cubic order
$$y(t)\ge\sqrt{2E_y}t-{1\over 3}\sqrt{2E_y}x(0)^2t^3.$$
Here, we have used the equation of motion to see that there is no quadratic
term in
$t$. The right hand side is positive for $t<\sqrt{3}/x(0)$. Therefore, we
have the following underestimate for the deceleration in $x$:
\begin{eqnarray}
  \int_0^{\sqrt{3}/x(0)} dt\, x(t) y(t)^2&>&
  x(0)\int_0^{\sqrt{3}/x(0)} dt\, y(t)^2\nonumber\\
  &>&2E_y x(0)\int_0^{\sqrt{3}/x(0)}dt\, \left(t-\frac 13 x(0)^2t^3\right)^2\nonumber\\
  &=&{16\sqrt{3}E_y\over 35 x(0)^2}.\nonumber
\end{eqnarray}
This is the deceleration for one half of a quasi-cycle of the $y$
motion. We have to sum the contributions of all the cycles. At first sight,
this sum (which should be thought of as an integral over $x$!) is convergent
because the contribution is proportional to $1/x^2$. But it is important to
note 
that this contribution comes from the interval $[0,t_1]$ and its length
is bounded by $\pi/x(0)$. Therefore, the average value is bounded from below
by
\begin{displaymath}
  {16\sqrt{3}E_y\over 35 \pi x(0)}
\end{displaymath}
for the cycle beginning at $t=0$. To sum the contribution from all values of
$x$ one has to sum a logarithmically divergent harmonic sum. This is in
accordance with the logarithmic divergence we expected from the heuristic
argument. 

Hence we found that, however large $\dot x$ and however small the
non-vanishing motion in $y$ is, the motion in $x$ will eventually be stopped
which is the contradiction we were looking for. Therefore, we can conclude that
the only motions that reach infinity are those along straight lines.

This argument has a straight forward generalization to more than two
variables: First, we can increase the number of degrees of freedom that play
the r\^ole of $y$ above:
\begin{displaymath}
  H=\frac 12\left(\dot x^2+\sum_i \dot y_i^2+x^2\sum_i y_i^2\right)
\end{displaymath}
The one dimensional harmonic oscillator is generalized to a multidimensional
harmonic oscillator of which the orbits are ellipses. A simple way to
see this is to introduce polar coordinates for the $y_i$ and observe that for
the radial coordinate there is a centrifugal term that prevents the oscillator
from approaching the origin. As the deceleration force on $x$ is proportional
to the square of the radial coordinate, the true, time-dependent force is
bounded from below from the constant force that corresponds to the minimal
distance to the origin in $y_i$-space.

In the case of vanishing angular momentum in $y_i$ space, the motion is
effectively one dimensional and we are back to the original toy model
discussed above.

A more ``democratic'' generalization is to consider the Hamiltonian
\begin{displaymath}
  H=\frac 12\left(\sum_i \dot x_i^2+\sum_{i\ne j} x_i^2x_j^2\right).
\end{displaymath}
Again, by symmetry, we can assume that the solution escapes along the positive
$x_1$ direction. Using the same argument as above we can conclude that 
\begin{displaymath}
  x_1\sim R=\sqrt{\sum_i x_i^2}
\end{displaymath}
and again, as the potential energy is bounded by the total energy
\begin{displaymath}
  x_i\sim \frac {\sqrt E}R
\end{displaymath}
for $i>1$. The equation of motion for $x_1$ is the same as for $x$ in the
first generalization above, whereas in
\begin{displaymath}
  \ddot x_i = -x_i\sum_{j\ne i}x_j^2
\end{displaymath}
the term with $j=1$ that corresponds to the equation of motion for $y_i$ above
dominates the other terms by four orders of $R$. Thus, in the limit of late
times and therefore large $R$, this second generalization reduces to the first.

\section{The full model}

Before starting to discuss how to extend the proof to the full model, let
apply a slight reformulation: It will be enough to consider the two particle
case as here one would already expect solutions that exchange transversal
momenta, i.e. that have different velocities in the in compared to the out
state.  Furthermore, we are not interested in the center of mass motion so we
can separate off the diagonal $U(1)$ and end up with a $su(2)$ model. Next, we
decompose everything in terms of the Pauli matrix basis and use a vector
notation for $su(2)$ indices. This turns the bosonic Lagrangian \ref{eq:MMLag}
into
\begin{eqnarray}
  L&=&\frac 12\dot{\vec X^2}\cdot \dot{\vec X^2} - \frac 14 \|\vec X^i \times
  \vec X^j\|^2\nonumber\\
  &=& \frac 12\dot{\vec X^2}\cdot \dot{\vec X^2} - \frac 14 \left(\sum_{i\ne
  j}\|\vec X^i\|^2\|\vec X^j\|^2 - (\vec X^i \cdot \vec X^j)^2\right)\nonumber
\end{eqnarray}

Note that for any given instant of time, the second term in the potential can
be made to vanish by a choice of coordinates in $\MR^9$ that diagonalizes the
symmetric matrix
\begin{displaymath}
  M^{ij} = \vec X^i\cdot \vec X^j
\end{displaymath}
In this choice of gauge, the model already looks similar to the second
generalization of the toy model in the previous section. Unfortunately, this
gauge is not obeyed by time evolution.

The toy model enjoys one important simplification compared to the full \MM:
There is no gauge freedom, the valleys are along the $x$- and $y$-axes whereas
the global $SO(9)\times SU(N)$ symmetry allows one to rotate the valleys in
any possible direction. The direction of the valleys is therefore dynamically
determined and one should use appropriate variables to deal with this
symmetry. Furthermore, one could imagine a solution in which the matrices
cannot asymptotically be brought to diagonal form because there is a
non-vanishing motion in the $SU(N)$ directions that would destroy any choice
of gauge at later times.

To approach these difficulties, let us first discuss another simplified model
that now contains all qualitative features of the full \MM; this allows one to
translate the argument directly to the full \MM, but we prefer to present the
approach in this model for notational simplicity.

The degrees of freedom are two two-vectors $\vec X_a(t)\in\MR^2$ with
$a=1,2$. The Lagrangian is given by (note the close similarity to the
Lagrangian of the $SU(2)$-\MM\ in the reformulation given above!)
\begin{displaymath}
  L=\frac 12\left(\|\dot{\vec X_1}{}\|^2+\|\dot{\vec
      X_2}{}\|^2\right)-\left(\vec X_1\times\vec X_2\right)^2.
\end{displaymath}
Note that also this model arises as a truncation of the full \MM\ if we define
$\vec e=(e_1,e_2)$ and use the ansatz
\begin{displaymath}
  X^i = \vec X_i\cdot \vec e
\end{displaymath}
for $i=1,2$ and take the other $X^i$'s to be zero.
Once again, we ask whether there are solutions such that
\begin{displaymath}
  R^2:=\|\vec X_1\|^2+\|\vec X_2\|^2
\end{displaymath}
goes to infinity for early and late times. The form of the potential energy
tells us that for large $\|\vec X_1\|$, say, the component of $\vec X_2$
that is perpendicular to $\vec X_1$ has to be small and oscillates with
approximate frequency $\|\vec X_1\|$. Nevertheless, it is not clear, that there
is an asymptotic direction for $\vec X_1$, it might rotate around the origin
forever. Therefore we cannot simply fix a gauge in which only one component of
$\vec X_2$ plays the r\^ ole of $y$ in the toy model.

To solve this problem, the first observation is that, just like the \MM, this
model not only has the obvious $SO(2)$ symmetry of vectors in $\MR^2$ that
parallels the $SU(N)$ symmetry of the \MM, but that it is invariant under
another $SO(2)$ that acts on the $a$ index of the $\vec X_a$, because the
potential is just the square of the determinant of the matrix $X_a{}^i$. A
parameterization that is adapted to the $SO(2)\times SO(2)$ symmetry is
\begin{displaymath}
  X_a{}^i=\pmatrix{\cos\alpha&\sin\alpha\cr -\sin\alpha&\cos\alpha\cr}
  \pmatrix{x&0\cr 0&y\cr}\pmatrix{\cos\beta&\sin\beta\cr
    -\sin\beta&\cos\beta\cr}.
\end{displaymath}
This parameterization is similar to the one used in \cite{SS} only that here the
matrices employing the diagonalization are time dependent.
If we rewrite the Lagrangian in these variables
\begin{displaymath}
  L=\frac 12\left( \dot x^2+\dot y^2+(x^2+y^2)(\dot\alpha^2+\dot\beta^2)
  +4xy\dot\alpha\dot\beta\right)-x^2y^2,
\end{displaymath}
we see that $\alpha$ and $\beta$ are cyclic variables and their conjugate
momenta
\begin{eqnarray}
p_\alpha&=&(x^2+y^2)\dot\alpha+2xy\dot \beta\nonumber\\
p_\beta&=&(x^2+y^2)\dot\beta+2xy\dot\alpha\nonumber
\end{eqnarray}
are integrals of motion. We solve these for $\dot\alpha$ and $\dot\beta$ and
use them to eliminate the $\alpha$ and $\beta$ dependence from the Lagrangian.

Then, we arrive at the equation of motion for the remaining degrees of freedom
\begin{equation}
  \label{eq:effeom}
  \ddot x=-2xy^2+{(p_\alpha^2+p_\beta^2)x(x^4+10x^2y^2+5y^4)-2p_\alpha p_\beta
    y(y^4+10x^2 y^2+5x^4)\over\left(x^2-y^2\right)^4}
\end{equation}
and another one with $x\leftrightarrow y$.
We use the same argument as before to show
\begin{displaymath}
  x>{R\over\sqrt{2}},\qquad y<{\sqrt{2}E\over R}
\end{displaymath}
which tells us that for large $R$ we have $x\sim R\gg y\sim 1/R$. While the
first 
term in \ref{eq:effeom}\ scales like $1/R$, the second term scales like
$1/R^3$ and can therefore be neglected for large $R$:
\begin{displaymath}
  \ddot x\approx-2xy^2+{(p_\alpha^2+p_\beta^2)x^5-p_\alpha p_\beta yx^4\over
    x^8}
\end{displaymath}
In the numerator, it is not possible that the coefficient of $x^5$ vanishes
while the coefficient of $yx^4$ is finite. Therefore, for large enough $R$
(depending on the two angular momenta), we can neglect the second term in the
numerator against the first. The fraction scales like $1/R^3$ and is always
strongly suppressed by the ``toy-model'' term.

Similarly, for large $R$, we have
\begin{displaymath}
  \ddot y\approx -2yx^2 +{5(p_\alpha^2+p_\beta^2)x^4y-2p_\alpha p_\beta
    x^5\over x^8}.
\end{displaymath}
The second term scales like $1/R^3$, whereas the first term scales like
$yR^2$. As the amplitudes of the oscillations in $y$ are of the order $1/R$
the second term will, for most of the time of one oscillation, be suppressed
by four orders of $R$ and can therefore be neglected.

We have argued that in both non-trivial equations of motion, the second terms
can be neglected and in the limit of large $R$ , we are left with the toy
model. In the previous section, we proved that there are no non-trivial
solution that can escape to infinity. Therefore there will be no scattering
solutions for this model, too.

We would now like to use the same approach like above for the $SO(2)\times
SO(2)$ model for the full $SO(9)\times SU(2)$ \MM\ by writing
\begin{displaymath}
  X^i_a = U^{ij}(t) \pmatrix{x(t)&&&\qquad\cr &y(t)&&\cr &&z(t)&\cr}_{jb}
  V_{ab}(t) 
\end{displaymath}
$X^i_a$ is a decomposition of $X^i$ in terms of the Pauli matrix basis. Here,
$i$ as before runs from 1 to 9 and $a$ runs from 1 to 3. $V$ is a matrix in
$SU(2)$ while $V$ is a matrix in the coset $SO(9)/SO(6)$ as an $SO(6)$ leaves
this form invariant. A simple counting of dimensions shows that this
parameterization is generically possible with singularities arising in cases of
the diagonal values coinciding. We can ignore this subtlety since it will not
play a r\^ole in what follows.

Again, the potential energy in this parameterization is given by
\begin{displaymath}
  V= x^2y^2 + y^2 z^2 + z^2 x^2
\end{displaymath}
and one would expect only these degrees of freedom to be ``physical'' whereas
the degrees of freedom encoded in $U$ and $V$ being gauge. Thus, the system
can be viewed as an extension of the solid body with dynamical moments of
inertia. 

Ideally, one would like to eliminate $U$ and $V$ including their time
derivatives in favor of the conserved momenta
\begin{displaymath}
  L^{ij} = \dot X^i_a X^j_a-X^i_a \dot X^j_a,\qquad M_{ab} = \dot X^i_a
  X^i_b-X^i_a \dot X^i_b
\end{displaymath}
($M_{ab}$ has to vanish due to the Gauss constraint) but this is not possible
in general for non-abelian gauge groups. One way of understanding is that some
components of the constraint $L^{ij}(X,\dot X)=\tilde L^{ij}$ are second class
in the sense of Dirac.

The full analysis has been carried out in \cite{DR} for the case relevant for
QCD, namely $SO(3)$ instead of $SO(9)$ that we will also restrict ourselves to
in the following. The $SO(9)$ case can be treated similarly. After rotation
of coordinates one can assume that
\begin{displaymath}
  k^k = \epsilon^{ijk} \tilde L^{ij}
\end{displaymath}
is of the form $k^i = (0,0,k)$. After performing the symmetry reduction one is
left in addition to $x$, $y$, and $z$ with an angle $\phi$ and its conjugate
momentum $p_\phi$ that is restricted to $p_\phi^2 \le k^2$. The reduced
Hamiltonian is given by
\begin{eqnarray}
  \label{redham}
  H &=& \frac 12\Big( \dot x^2 + \dot y^2 + \dot z^2 + x^2y^2 + y^2z^2 +
  z^2x^2\nonumber\\
&&\qquad+ D(x,y) p_\phi^2 + (k^2-p_\phi^2) (D(x,z) \sin^2\phi +
  D(y,z)\cos^2\phi ) \Big)
\end{eqnarray}
where we have introduced the abbreviation 
\begin{displaymath}
  D(a,b) = \frac{a^2 + b^2}{(a^2-b^2)^2}
\end{displaymath}
We immediately recognize the second generalization of the toy model as the
first line of \ref{redham}. Hence, we have to show that the additional terms
in the second line do not change the result, that there are no non-trivial
solutions for which any of $x$, $y$, or $z$ escapes to infinity. Again with
the help of permutation symmetry we only consider the case of $x\to\infty$. As
before, we find the following scaling behavior with $R=\sqrt{x^2+y^2+z^2}$:
\begin{displaymath}
  x\sim R,\qquad y\sim z\sim \frac 1R.
\end{displaymath}
For the new terms, we find
\begin{displaymath}
  D(x,y)\sim D(x,y)\sim \frac 1{x^2}\sim\frac 1{R^2}
\end{displaymath}
as well as
\begin{displaymath}
  p_\phi^2 \sim (k^2-p_\phi^2)\sim \sin^2\phi\sim \cos^2\phi\sim 1
\end{displaymath}
\epsfxsize = 0.3\hsize
Therefore, the leading terms in the potential are\EPSFIGURE{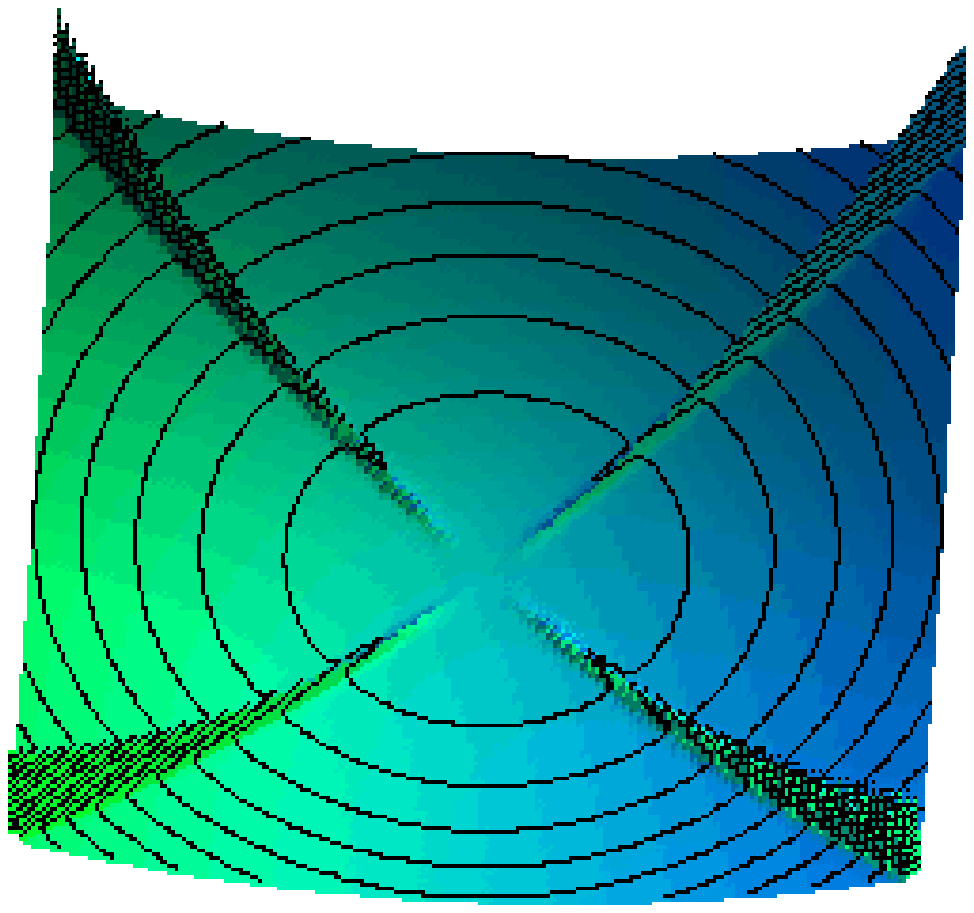}{The
  leading order potential $V_{leading}$}
\begin{displaymath}
  V_{\rm leading}=\frac 12 (x^2 y^2+ x^2 z^2 + \cos^2 \phi D(y,z))
\end{displaymath}
To get rid of the last term scaling analysis in terms of $R$ is not
enough. Rather, again, we should think of $x$ as being large and constant over
typical time-scales of the $(y,z)$ dynamics. After a last change of variables
to 
\begin{displaymath}
  y = r \sin\psi,\qquad z=r \cos\psi
\end{displaymath}
we find that for generic values the last term is strongly suppressed compared
to the harmonic oscillator term. Only if $y$ approaches $\pm z$ it acts as a
barrier and flips $\psi$ to $\pi/2-\psi$ while $r$ is not affected. This is a
discrete symmetry of the oscillator dynamics and therfore does not influence
the time dependence of $r$ which causes the deceleration of $x$ as in
the generalized toy model. 

Therefore, the analysis of the previous section carries over to the full \MM\
and we have proved that there are no non-trivial scattering solutions that
could describe transversal momentum transfer.

\section{Instanton tunneling}

In the previous section, we have seen that momentum transfer is not possible
classically in the matrix model. Thus, one cannot, quantum mechanically,
perform a fluctuation analysis around a classical solution. Nevertheless, it
might be that genuine non-perturbative quantum processes exist that
nevertheless have finite probabilities for processes with different in and out
states. In this case, momentum transfer should be viewed as a tunneling process
although there is no potential well that forbids the transition classically.

In this section, we are going to calculate a WKB transition probability for
such a process. For simplicity, we restrict our attention again to the toy
model. We perform a Wick rotation to the Euclidean domain that amounts to
flipping the sign of the potential energy. Instead of four valleys we have now
four basins that are not bounded from below. Those are separated by two
crossing wells along the coordinate axes that have their top at potential
energy zero.

In this energy landscape, one expects a motion that asymptotically comes in
from the positive $x$-axis then deviates into the north-east basin in a way
the down-hill force is balanced by the centrifugal force and then leaves
asymptotically along the positive $y$ axis. There is a simple argument that
such a solution has to exist\footnote{I am thankful to A. Smilga for an
  explanation of this point}: Take the particle to be on the line $x=y$ at
some instant of time with a velocity vector
\begin{equation}
  \label{eq:initv}
  (\dot x,\dot y) = (-v,v).
\end{equation}
For small values of $v$, the motion is not fast enough to go over the wall
along  the $y$-axis and the particle will end up in the north-east basin while
for large values of $v$ the motion will overshoot the wall along the $y$-axis
and the particle will roll down the north-west basin.

By continuity, there must be a value to which $v$ can be fine-tuned such that
it will just make it asymptotically to the top of the $y$-axis wall. Using
time reversal symmetry together with $x\leftrightarrow y$ one finds that for
this solution also comes asymptotically from the top of the $x$-axis.

Note that after fixing the critical value of $v$ also the asymptotic
velocities $v_\infty$ are determined by energy conservation. The only free
parameter is the point where the particle crosses the line $x=y$. This
remaining freedom comes from the scaling symmetry
\begin{displaymath}
  (x,y)\mapsto \lambda(x,y),\qquad t\mapsto \lambda^{-1}t,\qquad{\cal
  S}\mapsto \lambda^3{\cal S}
\end{displaymath}
of the model. Using this symmetry one finds a one to one correspondence between
asymptotic velocities and crossing points on $x=y$.

The only assumption we made about the orbit was \ref{eq:initv}. Dropping this,
it is highly unlikely that further solutions exist since for any other
direction of the initial velocity one would still have only one parameter to
tune (the magnitude of the velocity) but fine-tuning the final direction to be
along the top of the wall $x=0$ would destroy the fine-tuning for $t\to
-\infty$. Therefore the symmetric choice is the only possibility to achieve
$y\to0$ initially and $x\to 0$ for late times due to the instability of the
potential. 

Since there is not much hope to find the critical solution analytically, we
have used {\tt mathematica} to determine it
numerically.\EPSFIGURE{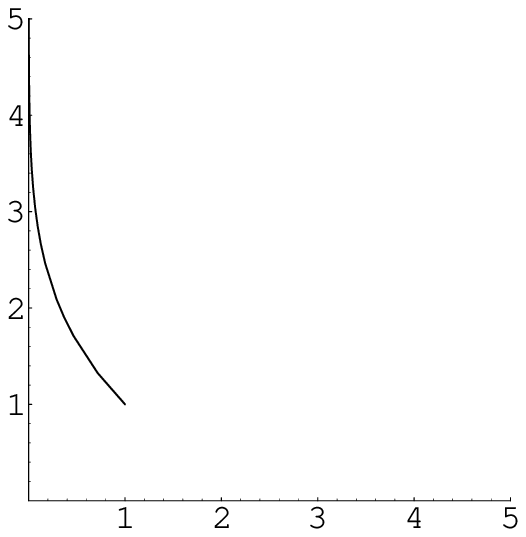}{The numerical solution with $v=v_{\rm
    crit}$}  We found
\begin{displaymath}
  v_{\rm crit} = 1.5099906769 x_0^2
\end{displaymath}
where $(x_0,x_0)$ is the initial point on the line $x=y$. The leading order of
the quantum mechanical probability for this amplitude is given by the classical
action of the orbit. As the initial and final velocities $v_\infty$ do not
vanish, the action is infinite. But so is the action for the trivial
solution. Thus we can calculate the difference between these two actions
yielding a relative probability:
\begin{displaymath}
  \frac{p({\rm trivial})}{p({\rm tunnel})}= e^{-({\cal S}_{\rm trivial} -
  {\cal S}_{\rm tunnel})},
\end{displaymath}
where the difference of the actions is given by
\begin{displaymath}
  {\cal S}_{\rm trivial} - {\cal S}_{\rm tunnel} = \int_{-\infty}^\infty\!\!\!
  dt\, \frac 12 (v_\infty^2-\dot x^2-\dot y^2 + x^2y^2)= 1.12 x_0^3
\end{displaymath}
where the integral again has been evaluated numerically. 

The integrand falls off exponentially with $|t|\to \infty$. Therefore, it is
no problem to perform the integration only for those values of $t$ for which
the numerics of the solution are stable (given the instability of the
euclidian potential).

\section{Conclusions}

In this note, we have shown that there are no non-trivial scattering solutions
in the \MM. Therefore it is classically impossible to study processes in which
transversal momentum is transferred between the two partons. This is quite
different compared to what is known about the 1+1-dimensional case of
M(atrix)-String-Theory. There, such solutions have been constructed. 

As the \MM\ and M(atrix)-String-Theory are only equivalent (via T-duality) in
the $N\to\infty$ limit, these findings cast further doubt on the usefulness of
the finite $N$ version of the \MM-conjecture of \cite{Su}. Rather it seems
there are quite strong dynamical restrictions on the super-gravity kinematics
that this model is able to describe.

The result can also be formulated reversely: Among all classical orbits of the
\MM\ equations of motion only a zero-set (namely the trivial solutions) are
scattering solutions whereas all possible other ones are bound solutions that
cannot get away from the stadium. 

On the other hand, this is quite a welcome behavior for the super-membrane
interpretation\cite{dWHN} of the \MM: There, what are scattering solutions
from the D0-particle perspective are degenerate solutions for which for early
and late times all the energy of the membrane is in the infinite growth of a
single mode of the world-volume coordinates. The bound solutions are more like
what one would expect of a fluctuating membrane.

\acknowledgments
I have benefited very much from various discussions with many people. Among
them are H.~Nicolai, J.~Hoppe, A.~Rendall, A.~Smilga, and
A. Waldron. Furthermore, I thank the Max-Planck-Society for financial support.

\bibliography{memscat}

\end{document}